\documentclass{article}
\usepackage{graphicx} 

\def\beq{\begin{equation}}
\def\eeq{\end{equation}}

\def\m#1{$#1$}

\usepackage{amssymb}
\def\beqs{\begin{eqnarray}}
\def\eeqs{\end{eqnarray}}
\def\[{\left[}
\def\]{\right]}
\def\({\left(}
\def\){\right)}

\def\eps{\epsilon}

\def\pmb#1{{\mathbf #1}}
\def\text#1{#1}
\def\textit#1{{\it #1}}
\begin{document}

\centerline{\bf A Relativistic Wave Equation  for the Skyrmion }

\begin{center}{\sf S. G. Rajeev}\\
Department of Physics and Astronomy\\ 
Department of Mathematics\\
University of Rochester\\
  Rochester NY 14627
\end{center}


\centerline{\bf Abstract}
    
    We propose a relativistically invariant wave equation for the  Skyrme soliton. It is a differential equation on the space $R^{1,3}\times S^3$ which is invariant under the Lorentz group and isospin. The internal variable valued in $SU(2)\approx S^3$ describes the orientation of the soliton. The mass of a particle of spin and isospin both  equal to $j={1\over 2},{3\over 2}\cdots$  is predicted to be $M=m\sqrt{1+K_2j(j+1)\over 1+K_1j(j+1)}$ which agrees with the known spectrum for  low angular momentum. The iso-scalar magnetic moment is predicted to be $-{K_1\over 4m}{\mathbf \Sigma}$, where ${\mathbf \Sigma}$ is the  spin.

\pagebreak
\section{ Introduction}

Relativistic wave equations for extended objects, as opposed to fundamental particles, are hard to find. Perhaps the most well-known such theories of this type are string and membrane theories. Yet we know that many relativistic field theories have soliton solutions. There ought to be relativistic wave equations that describe their motion. 
 The most physically interesting of these is the Skyrme soliton\cite{Skyrme,Syracuse,Witten}  which describes the baryon (such a a proton) as a  topologically non-trivial configuration  of pi mesons. 

There is a quantum theory\cite{BalTasi}  of the collective excitations of this solution, which predicts a mass spectrum
\beq
M=m\left[1+{1\over 2}Kj(j+1)\right]
\eeq
where $K$ is a kind of moment of inertia. If the number of colors $N_c$ of the underlying fundamental theory of strong interactions (QCD) is odd, the angular momentum must take on half-odd-integer values
\beq
j={1\over 2},{3\over 2}\cdots
\eeq
If $N_c$ is even,  the  values of $j$ are integers ${1,2,\cdots}$. In either case, each such state also carries an isospin $I=j$. 
Of course, once interactions with the pion waves are allowed, the higher energy states ought to be unstable against decay to the ground state, the nucleon.

Since the soliton arises in a relativistic theory, we should expect that there is some relativistically invariant wave equation that describes all of these excitations. In the usual theory of collective excitations,  the soliton is thought of as a heavy particle, with only small changes in energy allowed from its ground states. Thus the soliton must have small kinetic energy (compared to rest mass). Also, it must have  small angular momentum, so that the velocity of rotation at any point in its interior  is small compared to the velocity of light.   As the rotation of the soliton increases, it will change its shape (due to centrifugal forces) and we should not expect the above formula for rest mass to hold for large $J$. Nevertheless, it is interesting to ask if there is a relativistic wave equation that predicts   the above spectrum for small angular momentum.

There were many attempts to find physically interesting relativistic wave equations for particles with internal degrees of freedom\cite{Harish-Chandra}. They were motivated largely by the  desire to find a relativistic wave equation with positive energy spectrum. Although none of them turned  to be of use in physics, these investigations led to profound developments by Harish-Chandra  in the representation theory of semi-simple groups\cite{HarishHistory}. Our investigation can  be considered a continuation of this theme. We will not insist on having positive energy though, as it has turned out that this is not a physically realistic restriction.  Although we do not address this issue here, we expect that the problem of negative energies is solved by second quantization, as with all the other successful wave equations.

 In 2+1 dimensions, Jackiw and Nair\cite{JackiwNair}  have found relativistic wave equations for particles with fractional spin. 

Our wave equation is a differential equation for a complex-valued function on $R^{1,3}\times S^3$. The internal degree of freedom $S^3$ leads to spin as well as isospin. There could be an entirely different application of this idea to Kaluza-Klein theories, in which $S^3$ represents compactified  dimensions of space-time. The spectrum it predicts does not, however,  look like that of elementary particles we know currently.

Skyrmions have recently  reincarnated as collective excitations of the Bose-Einstein condensate. However, at the moment we don't see any reason why their spectrum must be relativistically invariant.

\section{The Non-Relativistic Wave Equation}
In the Skyrme model the instantaneous wave function of the baryon is a function of the space, time, and $g\in $SU(2) which describes its orientation in internal (isospin)  space. If the baryon is a boson  (resp. fermion), the
wave function is invariant ( resp. changes sign) under reflections in SU(2). Define the space of  spinors\footnote{A more intrinsic point of view is that $V_\pm$  is the space of sections of a complex line bundle on SO(3)=SU(2)/$Z_2$. There are exactly two such line bundles: the trivial one who sections are $V_+$ defined above and the non-trivial one which has $V_-$ as sections. We prefer the definition given here as it is more explicit.
}

\beq
V_{\pm }\equiv L^2{}_{\pm }(\text{SU}(2))=\{\chi :\text{SU}(2)\to C|\chi (-g)=\pm \chi (g)\}
\eeq

$V_{\pm }$  carries a repesentation of SU(2)$\times $SU(2), the left and right actions of SU(2) on itself. The wavefunction depends on space-time and takes values in $V_\pm$
\beq
\psi :R^{1,3}\to V_{\pm }.
\eeq
Thus $V_\pm$ plays a role analogous to the space of spinors in Dirac's theory. The rotation group of space (generating angular momentum) acts on  the left:
\beq
(x,g)\to  (R(h)x,h g), h\in \text{SU}(2).
\eeq
where  R:SU(2)$\to $ SO(3) is the usual  homomorphism into the
rotation group. Isospin acts on the other side of  $g$ and leaves the position $x$ invariant:
\beq
  (x,g)\to  \left(x, g h^{-1}\right).
  \eeq
 Using the Peter-Weyl theorem of harmonic analysis\cite{Weyl}, we can expand this wave function in terms of irreducible representations of SU(2)$\times$ SU(2):
 \beq
 \psi(x,g)=\sum_{j\in N_\pm}\psi_{jmm'}d^j_{mm'}(g)
 \eeq
We denote by \m{N_\pm} the sets $\{1,2,\cdots\}$ and  $\{{1\over 2},{3\over 2}\cdots\}$ respectively.

   takes  half-integer values for $V_-$ and integer values for $V_+$. Also, $m,m'=-j,-j+1,\cdots j$ and $d^j_{mm'}(g)$ are the representation matrices for the spin $j$ representation of SU(2). Thus we see that the isospin and spin have equal magnitudes  $j(j+1)$ in each irreducible component.
  
  The rotational energy  of a baryon at rest is given  in terms  of the Laplace  operator on \text{SU}(2) as    $\frac{m}{2}K \Delta _g$, where  $m$ is the mass of the lightest baryon and  $K$ is  proprtional to the  moment of inertia of the soliton.  In the units we use, $\hbar =c=1$ and $K$ 
is a dimensionless quantity. Recall that in terms of the generators of the left and right action of SU(2) on itself, 
\beq
\Delta _g=\pmb{\Sigma }^2{}_L=\pmb{\Sigma }_R{}^2
\eeq
Thus,  in each irreducible component the rotational energy has eigenvalues
\beq
{m\over 2}Kj(j+1).
\eeq
  
  The Schrodinger equation  for the free Skyrmion is then
  \beq
  -i \frac{\partial \psi }{\partial t} =\left[\frac{1}{2m}\Delta +\frac{m}{2}K\text{  }\Delta _g\right]\psi \label{Schr}
  \eeq
  where $\Delta=-\nabla^2$ is the Laplace operator on space $R^3$.
 This equation has too much symmetry: it is invariant under separate left and right actions of SU(2) on $g$  as well as under the SO(3) action on  $x$. In other words it conserves the orbital angular momentum and spin separately. This is also true of the non-relativistic  free (Pauli) wave equation for the electron.  The correct  relativistic wave equation should break this unwanted symmetry so that only the sum of orbital and spin angular momentum is conserved, as for the Dirac equation.
  
 \section{The SL(2,C) Representation on $V_{\pm}$}
  
  The universal cover of the Lorentz group, SL(2,C), is the complexification of SU(2). Hence, a  representation of the Lie algebra su(2) on any complex vector space can be extended to a representation of sl(2,C).To illustrate this idea,  if we were to do this extension  to the defining representation of SU(2), we will get the  two dimensional (Pauli) spinor  representation of SL(2,C).
  
   Explicitly, if the su(2) representation is given by hermitean operators 
 $\pmb{\Sigma }$, we will have
 \beq
 [\Sigma_1,\Sigma_2]=\sqrt{-1}\Sigma_3, \quad  [\Sigma_2,\Sigma_3]=\sqrt{-1}\Sigma_1,\quad  [\Sigma_3,\Sigma_1]=\sqrt{-1}\Sigma_2
 \eeq
 or more succinctly
 \beq
 [\Sigma_i,\Sigma_j]=\sqrt{-1}\eps_{ijk}\Sigma_k
 \eeq
The six generators of the sl(2,C) action are then given by 
\beq
\Sigma_{01}=-i\Sigma_1,\quad \Sigma_{02}=-i\Sigma_2,\quad \Sigma_{03}=-i\Sigma_3 
\eeq
for the boost and 
\beq
\Sigma_{23}=\Sigma_1,\quad \Sigma_{31}=\Sigma_2,\quad \Sigma_{12}=\Sigma_3 
\eeq
  for the rotation generators. They satisfy the sl(2,C) relations, as can be directly verified:
 \beq
 [\Sigma_{\mu\nu},\Sigma_{\rho\sigma}]=\sqrt{-1}\left[
 \eta_{\mu\rho}\Sigma_{\nu\sigma}- \eta_{\nu\rho}\Sigma_{\mu\sigma} -\eta_{\mu\sigma}\Sigma_{\nu\rho} +\eta_{\nu\sigma}\Sigma_{\mu\rho}
 \right]
 \eeq
  In our case this is an infinite dimensional non-unitary representation of the Lorentz group which commutes with isospin; i.e.,  the right action of SU(2).
 \section{Relativistic Invariance}
 
 Let us see how the above  wave equation (\ref{Schr}) can be turned into a relativistically invariant one. The obvious guess  is the Klein-Gordon type equation on functions  $\psi:R^{1,3}\to V_{\pm }$,
\beq
   \left[\partial _{\mu }\partial ^{\mu }+m^2 K\text{  }\Delta _g+m^2\right]\psi =0
  \eeq
  This gives the  mass shell condition:
  \beq
  p_{\mu }p^{\mu }=m^2+m^2\text{  }K j(j+1),\quad j\in N_\pm.
  \eeq
As is common in relativistic wave equations, the energy can be positive or negative:
  \beq
  E=\pm \sqrt{\pmb{p}^2+m^2+ m^2\text{  }K j(j+1)}
  \eeq
  The negative energy solutions have to reinterpreted in terms of anti-particles. We postpone this issue for now and concentrate on the positive root. 
  
  In the non-relativistic limit, kinetic energies due to translation and and rotation are both small compared to the rest-mass:
\beq
E=\pm \sqrt{{\mathbf p}^2+m^2+ m^2 K j(j+1)}\simeq m +\frac{m}{2}K\text{  }j(j+1)+\frac{{\mathbf p}^2}{2m}
\eeq  
thus recovering the earlier formula, except for the addition of the rest mass to the energy
  
 But again this equation has too much invariance. It is invariant under
  $\text{SO}(1, 3)\times \text{SU}_L(2)\times \text{SU}_R(2)$
 The invariance should  under $\text{SO}(1, 3)\times \text{SU}_R(2)$. Unlike in the non-relativistic case, this error is not forgivable: only  orbital plus spin angular momentum can be conserved in a relativistic theory, not each separately. 
  There must be a  simultaneous action of SO(1,3) on space-time and SL(2,C) on $S^3\equiv \text{SU}(2)$  which ties Lorentz transformations to the left action on SU(2). 
  
  To break the symmetry,  recall  the Pauli-Lubansky vector 
  \beq
  W_{\mu }= \frac{1}{2} \epsilon _{\mu \nu \rho \sigma }p^{\nu }\Sigma ^{\rho \sigma }
  \eeq
  This is a Lorentz-vector only under the {\em simultaneous}  action on space-time and $V_\pm$. So we can add this to the wave equation to get an equation with the right symmetry:
  \beq
  \left[-p_{\mu }p^{\mu }+K_1 W_{\mu }W^{\mu }+m^2 K_2\text{  }\Delta _g+m^2\right]\psi=0. \label{RelWaveEqn}
  \eeq
  The dimensionless constants $K_{1,2}$   will be determined by physical properties of the soliton. In the rest-frame  $p\simeq (M,0,0,0)$ and $W\simeq (0,-M)$  and  
  \beq
W_{\mu } W^{\mu } \simeq  -M^2j(j+1), 
\eeq
since $\pmb{\Sigma}$  generates the left action  of the SU(2) Lie algebra on SU(2). So, the rest energy spectrum is now
\beq
-M^2-M^2 K_1 j(j+1)+m^2 K_2 j(j+1)+m^2=0
\eeq
  or
  \beq
  M=\pm m \sqrt{\frac{1+K_2 j(j+1)}{1+K_1 j(j+1)}}.
  \eeq
  Of course, each eigen-space with fixed $j$  has degenarcy $2j+1$ as it carries also a representation of \m{SU(2)_R} with iso-spin $j$.
For small isospin and spin, 
\beq
M \simeq  m + \frac{m}{2} K j(j+1),\quad K=K_2-K_1.
\eeq  
 in agreement with the non-relativistic spectrum. 
 
 If we wan the mass to remain finite for large $j$, we must impose
 \beq
 K_2>K_1>0.
 \eeq
In this case, as angular momentum grows, the energy tends to a constant:
\beq
E\to m\sqrt{\frac{K_2}{K_1}}>m
\eeq  
  This can be interpreted as a deformation of the soliton due to rotation that increases its moment of inertia. In this picture, for large angular momentum the moment of inertia tends to infinity, and the angular velocity tends to a constant. 
  
  It is quite possible that for large enough angular momentum the energy of the skyrmion is complex: it decays by emitting pions\cite{BKS}

  We need to know one extra piece of information  about the Skyrmion,  in addition to  the low energy spectrum,  to determine the two parameters \m{K_{1,2}}. The magnetic moment gives this piece.

  \section{ Electromagnetic Coupling}
  Recall that in the standard model of elementary particles, the electric charge of a  hadron  is isospin component plus half the baryon number:
  \beq
  Q=I_3+{B\over 2}
  \eeq
  The  isovector part seems to be more subtle in the Skyrme model\cite{Manohar}

E. Jenkins and A. V. Manohar Phys. Lett. B335, 452 (1994)  
A. V. Manohar,  Les Houches  Lectures 1997
  
  So the minimal coupling rule
  \beq
  p_{\mu }\to \pi _{\mu }=p_{\mu }-eA_{\mu }
\eeq  
applied to the wave equation ought to tell us about magnetic moment of the Skyrmion. It can  fix the extra  constant $K_1$. We get 
\beq
\left[-\pi _{\mu }\pi ^{\mu }+K_1 \omega _{\mu }\omega ^{\mu }+m^2 K_2\text{  }\Delta _g+m^2\right]\psi=0,\quad  \omega _{\mu }= \frac{1}{2}\text{  }\epsilon _{\mu \nu \rho \sigma }\pi ^{\nu }\Sigma ^{\rho \sigma }.\label{EM}
\eeq

If we take the non-relativistic limit in the presence of a weak constant magnetic field, 
\beq
\omega _0={\mathbf\pi} .\pmb{\Sigma },\quad  {\mathbf\omega }\approx -\pi _0\pmb{\Sigma}
\eeq
\beqs
\omega _{\mu }\omega ^{\mu }&\approx& (\pmb{\pi .\Sigma }\pmb{)}^2-\pi _0{}^2\pmb{\Sigma }^2\cr
&=&
-\pi _0{}^2\pmb{\Sigma }^2+\frac{1}{4}\left[\pi _i,\pi _j\right]\left[\Sigma _i,\Sigma _j\right]+\frac{1}{4}\left[\pi _i,\pi _j\right]_+\left[\Sigma _i,\Sigma _j\right]_+\text{                                                     }\cr
&=&-\frac{1}{2}\pmb{\Sigma .B}\pmb{+}\frac{1}{4}\left[\pi _i,\pi _j\right]_+\left[\Sigma _i,\Sigma _j\right]_+-\pi _0{}^2\pmb{\Sigma }^2
\eeqs
So that the magnetic energy in the NR approximation is
$ -\frac{K_1}{4m} \pmb{\Sigma .B} $. This gives the gyromagnetic ratio
\beq
\gamma =-\frac{K_1}{4m}.
\eeq
\section{The Complex Analytic Point of View}

There is another way to think of the representation of SL(2,C) on $V_\pm$.There is a representation of SL(2,C)$\times$ SU(2) on complex analytic functions on SL(2,C:
\beq
\phi(\gamma)\mapsto \phi(\lambda\gamma h^{-1}),\quad \lambda\in {\rm SL(2,C)}, h\in {\rm SU(2)}.
\eeq
Now, every analytic function $\phi:{\rm SL(2,C)}\to C$ can be obtained uniquely\cite{DriverGross,Hall} as the continuation of a  function $\psi$ on SU(2).  It is clear that even ( resp. odd) go over to even (odd functions) functions:
\beq
\psi(-g)=\pm \psi(g)\Leftrightarrow \phi(-\gamma)=\pm \phi(\gamma)
\eeq
and that this condition is preserved by the action of SL(2,C)$\times$ SU(2).
Thus we can identify $V_\pm$ with the space of analytic functions satisfying the above condition.  Thus we get a representation of SL(2,C) on $V_\pm$.

This is analogous to the way that the space of functions on the real line can be identified with holomorphic functions on the complex plane: the well-known Bargmann correspondence useful in studying coherent states\cite{Bargmann}.
Indeed SL(2,C) is diffeomorphic to the co-tangent space of $S^3$. Thus we are thinking of wavefunctions as analytic functions on the phase space instead of as complex-valued functions on the configuration space, just as with  coherent states.

There is a unique (up to multiplication) Riemannian metric that is invariant under the SL(2,C)$\times$ SU(2) action. This is not the familiar metric on a simple Lie group that is invariant under the left and right actions: on a non-compact group like SL(2,C) such a metric would not be positive. If we take the positive quadratic form in left-invariant vector fields, we will get instead  a Riemann metric invariant under SL(2,C$)_L\times$ SU(2$)_R$. The Laplacian of this metric is, up to a constant multiple, the operator $Delta_g$ on $V_\pm$ under the above identification.

\section{Some Red Herrings and Future Directions}

Is it possible to have an action of SL(2,C)$\times$ SU(2) on $S^3$ without the above analytic continuation to SL(2,C)? There are indeed ways that SL(2,C) can act on $S^3$ itself. They have all been classified\cite{Asoh}, but none commute with the right action of SU(2). Being not  invariant under isospin, they are not useful to us.

In the absence of the term with $K_2$, the mass of particles would have decreased with angular momentum, a physically unacceptable answer. This is typical of the wave equations studied by Bhabha and Harish-Chandra. 

I was unable to find a first order wave equation analogous to the Dirac  equation for our case. However, Feynman and Gell-Mann pointed out \cite{FeynmanGellMann}years ago that  the Dirac equation can be recast as a second order equation for a Pauli spinor (which has the half the components of a Dirac spinor, thereby compensating for the increase in order of the equation). Our equation could be viewed as a generalization to higher spins of this form of Dirac's equation.

We would like to understand the isovector part of magnetic moments as well. Also coupling to pions and the nature of chiral symmetry need to be understood. Although our wave equation has a term which is a fourth order differential operator, in the space-time co-ordinates it is still second order. It would be interesting to determine whether  the Cauchy problem is well-posed  for this equation.

\section{Acknowledgements}

I thank L. Gross, B. Hall   and A. Jordan for discussions. This work was supported in part by the Department of Energy   under the  contract number  DE-FG02-91ER40685.

 \pagebreak
 
\end{document}